\documentclass[%
 aip,
 jcp,%
 amsmath,amssymb,
preprint,%
]{revtex4-1}

\usepackage{graphicx}% Include figure files
\usepackage{dcolumn}% Align table columns on decimal point
\usepackage{bm}% bold math
\usepackage[version=3]{mhchem}

\usepackage{hyperref}

\usepackage{epstopdf} % should this be removed by the journal??

\begin{document}

\preprint{AIP/123-QED}

\title[Improved Transverse Confinement in a Zeeman Decelerator]{Getting a Grip on the Transverse Motion in a Zeeman Decelerator}

\author{Katrin Dulitz}
\affiliation{Department of Chemistry, University of Oxford, Chemistry Research Laboratory, 12 Mansfield Road, Oxford, OX1 3TA, United Kingdom}
\author{Michael Motsch}
\affiliation{Laboratorium f\"{u}r Physikalische Chemie, ETH Z\"{u}rich, Wolfgang-Pauli-Stra\ss{}e 10, 8093 Z\"{u}rich, Switzerland}
\author{Nicolas Vanhaecke}
\affiliation{Fritz-Haber-Institut der Max-Planck-Gesellschaft, Faradayweg 4--6, 14195 Berlin, Germany}
\affiliation{Laboratoire Aim\'{e} Cotton, CNRS, Universit\'{e} Paris-Sud, ENS Cachan, 91405 Orsay, France}
\author{Timothy P. Softley}
\email{tim.softley@chem.ox.ac.uk.}
\affiliation{Department of Chemistry, University of Oxford, Chemistry Research Laboratory, 12 Mansfield Road, Oxford, OX1 3TA, United Kingdom}

\date{\today}% It is always \today, today,

\begin{abstract}

Zeeman deceleration is an experimental technique in which inhomogeneous, time-dependent magnetic fields generated inside an array of solenoid coils are used to manipulate the velocity of a supersonic beam. A 12-stage Zeeman decelerator has been built and characterized using hydrogen atoms as a test system. The instrument has several original features including the possibility to replace each deceleration coil individually. In this article, we give a detailed description of the experimental setup, and illustrate its performance. We demonstrate that the overall acceptance in a Zeeman decelerator can be significantly increased with only minor changes to the setup itself. This is achieved by applying a rather low, anti-parallel magnetic field in one of the solenoid coils that forms a temporally varying quadrupole field, and improves particle confinement in the transverse direction. The results are reproduced by three-dimensional numerical particle trajectory simulations thus allowing for a rigorous analysis of the experimental data. The findings suggest the use of a modified coil configuration to improve transverse focusing during the deceleration process.

\end{abstract}

\pacs{Valid PACS appear here}% PACS, the Physics and Astronomy

\maketitle

\section{\label{sec:introduction} Introduction}

The generation and study of cold molecules is a vibrant research area with applications both in the chemical and physical sciences, for example, in cold chemistry \cite{ Willitsch2008, Henson2012} and cold collision dynamics \cite{Gilijamse2006a, Sawyer2008a, Kirste2012}, high-resolution spectroscopy \cite{Veldhoven2004, Hudson2006} or tests of fundamental physical theories \cite{Bethlem2008a, Hudson2011, DiSciacca2012, Campbell2013}. There have been a number of review articles summarizing the advances in this field \cite{Doyle2004, Schnell2009, Bell2009, Carr2009}.

% Research in the Softley group is directed towards the study of ion-molecule reactions at milliKelvin temperatures. Besides their importance in the context of astrophysics, these reactions are valuable probes of chemical reactivity in a regime that is dominated by quantum effects.
% Previous experiments  focused on reactions between Coulomb-crystallized \ce{Ca+} ions in a radiofrequency ion trap and polar molecules from a bent electric quadrupole guide \cite{Willitsch2008, Bell2009a}. Due to the broad energy range covered by the quadrupole guide, these reactions were not quantum-state selected and could only be observed at temperatures merely approaching the cold regime ($T\approx$~4~K). Nevertheless this was an early example of how two sources of cold species could be combined to study cold chemistry, and it illustrated the extra-ordinary sensitivity that is achievable using trapped ions as a collision target.
Chemical reactions involving {\it radicals} are of particular relevance for cold chemistry and the astrophysical medium, since they are fast, often barrierless, processes giving rise to rate constants on the order of 10$^{-10}$~cm$^{3}$~molecule$^{-1}$~s$^{-1}$, even at interstellar temperatures of a few Kelvin \cite{Anicich2003, Snow2008}. The large magnetic moments of these paramagnetic species provide the opportunity to control their translational motion using inhomogeneous magnetic fields.

In this paper, we report on the design and performance of a Zeeman decelerator that has recently been constructed and put into operation in Oxford. The manipulation of paramagnetic atoms and molecules in a Zeeman decelerator produces milliKelvin-cold, velocity-tunable beams of particles in specific internal quantum states. If combined with an ion trap, this technique should allow for the study of cold ion-radical processes as a function of collision energy, for example, the reaction between decelerated hydrogen atoms and sympathetically cooled \ce{CO2+} ions. The quantum-state selectivity of the deceleration process will greatly simplify the interpretation of kinetic data, because only certain reaction pathways are possible.

All supersonic beam deceleration techniques make use of the interaction between neutral particles and time-dependent, inhomogeneous electromagnetic fields to remove kinetic energy from a supersonic jet \cite{Hogan2011, van2012, Narevicius2012}. Depending on the type of electromagnetic interaction, these slowing methods can be further subdivided into Stark deceleration, Rydberg Stark deceleration, Zeeman deceleration and optical deceleration. This class of methodologies  was pioneered by the Meijer group, which was the first to decelerate beams of polar molecules, such as metastable \ce{CO} and \ce{ND3}, by rapidly switching electric fields of several tens of kV/cm in a Stark decelerator \cite{Bethlem1999, Bethlem2000a}. Due to their huge electric dipole moments, atoms and molecules in Rydberg states require much lower electric fields for deceleration \cite{Procter2003, Vliegen2004}. The deceleration of neutral molecules in an optical Stark decelerator is based on the optical dipole force induced by very intense pulsed laser fields \cite{Fulton2004}. Recently, a technique relying on the centrifugal force on a rotating disk has been developed to decelerate a continuous molecular beam \cite{Chervenko2014}.

Zeeman deceleration, originally developed by the groups of Merkt \cite{Vanhaecke2007a} and Raizen \cite{Narevicius2008}, makes it possible to manipulate the motion of open-shell atoms and molecules that cannot be addressed with electric fields. To date, this technique has been successfully used to decelerate supersonic beams of \ce{H} \cite{Vanhaecke2007a, Hogan2007}, \ce{D} \cite{Hogan2007}, metastable \ce{Ne} \cite{Narevicius2008, Wiederkehr2011}, metastable \ce{Ar} \cite{Trimeche2011}, \ce{O2} \cite{Narevicius2008a, Wiederkehr2012} and \ce{CH3} \cite{Momose2013}.
Like other supersonic beam deceleration experiments, Zeeman deceleration starts with a supersonic expansion from a high-pressure gas reservoir into a vacuum. While the particles' internal degrees of freedom are adiabatically cooled, the beam exits the expansion region with a high forward velocity of several hundred meters per second, depending on the carrier gas used. Zeeman deceleration relies on the influence of inhomogeneous magnetic fields on the motion of paramagnetic particles. These magnetic fields are produced by successively pulsing high currents through an array of solenoid coils. Upon approaching the maximum magnetic field in the center of a solenoid coil, the Zeeman energy of particles in low-field-seeking quantum states is increased, while their kinetic energy is decreased. To obtain a net loss of kinetic energy, the magnetic field is rapidly switched off before the particles reach the negative slope of the solenoid magnetic field. This process is repeated in subsequent coils until a desired final velocity is reached. The amount of kinetic energy removed depends on the magnetic moment of a quantum state, thus providing a means to manipulate the motion of particles in specific quantum states. Phase-space stability ensures that a bunch of particles is kept together throughout the deceleration sequence.

Research on cold chemistry and cold collisions would benefit from high number densities, as these would make experiments easier and less time-consuming to carry out. However, due to several different loss processes, the particle densities of decelerated beams are rather low, typically on the order of 10$^{9}$~molecules~cm$^{-3}$ or less \cite{van2009}.
The principal factors limiting the density are two-fold. First, a rapid reversal of the magnetic or electric field direction can cause Majorana transitions to high-field-seeking quantum states that are not decelerated \cite{Vanhaecke2007a, Hogan2008a, Wall2010, Meek2011a}. For Zeeman deceleration, these losses can be suppressed by temporally overlapping the current pulses of adjacent coils \cite{Hogan2008a}. In a Stark decelerator, transitions to high-field-seeking states were circumvented by switching the decelerator electrodes between a high and a low electric field configuration, as opposed to turning off the electric field of an electrode pair after each deceleration step \cite{Wall2010}. Nonadiabatic transitions in a chip-based Stark decelerator were reduced by applying a static offset magnetic field perpendicular to the electric field such that the energetic splitting between the hyperfine states at zero electric field was increased \cite{Meek2011a}.

Second, longitudinal and transverse effects can reduce the phase space acceptance of a decelerator. Van de Meerakker and co-workers showed that an interplay between the longitudinal and transverse motion set limits to the efficiency of Stark deceleration \cite{van2006}. Transverse effects, such as overfocusing, defocusing and insufficient transverse confinement, represent other important loss processes \cite{van2006, Wiederkehr2010}. Several strategies have been proposed and pursued to increase the particle flux. The transverse confinement was enhanced using novel types of decelerators that rely on co-moving magnetic or electric traps to confine and decelerate the particles \cite{Osterwalder2010, Trimeche2011, LavertOfir2011a, Hogan2012a}. There has also been a proposal for a quadrupole-guiding Stark decelerator, in which the longitudinal and the transverse motion are decoupled through the use of additional electrodes for transverse focusing \cite{Sawyer2008}. Other approaches involve the use of higher-order modes for deceleration or the application of optimized pulse sequences based on evolutionary algorithms \cite{van2006, Wiederkehr2010}. There are also other losses related to mechanical heating \cite{Meek2011a} or the imperfect mechanical alignment of the deceleration stages \cite{QuinteroPerez2013}, but those are specific to each decelerator setup.

In this article, we give a detailed description of the experimental setup of the new Oxford Zeeman decelerator, comprising a number of advancements with respect to existing experiments. 
Since the Zeeman deceleration of hydrogen atoms has been successfully demonstrated in previous studies \cite{Vanhaecke2007a, Hogan2007, Hogan2008a, Hogan2008b}, we have chosen this atomic species in order to characterize the performance of our apparatus. We give evidence that the transverse confinement inside a Zeeman decelerator is increased by applying a low, anti-parallel magnetic field to one of the coils. Our results are compared to three-dimensional numerical particle trajectory simulations, which also facilitate a more general understanding of the experimental findings.

\section{\label{sec:experimental} Experimental}

\subsection{\label{sec:generalsetup} General Setup}

\begin{figure}
\includegraphics{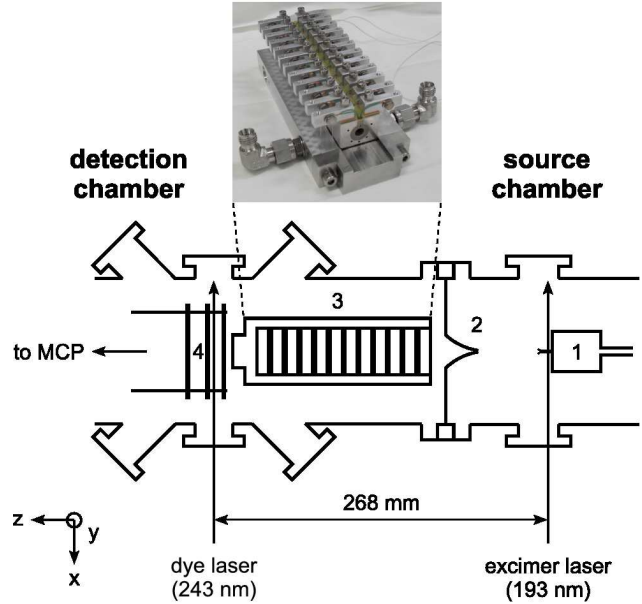}
\caption{\label{fig:sketchchamber} Sketch of the experimental setup, not to scale. 1: Pulsed valve, 2: Skimmer, 3: Zeeman decelerator, 4: Extraction plates. A photograph of the Zeeman decelerator is shown as an inset.}
\end{figure}

A schematic illustration of the experimental setup, along with a picture of the Zeeman decelerator, is shown in Figure~\ref{fig:sketchchamber}. The molecular beam apparatus consists of two chambers: a source chamber for the generation of a supersonic hydrogen atom beam, and a detection chamber containing both the Zeeman decelerator and a time-of-flight (TOF) detection system. The chambers are separated by a skimmer (Beam Dynamics, 2.0~mm orifice diameter), and differential pumping is achieved using two turbo pumps (\mbox{690~l/s}, Turbovac SL700, Oerlikon Leybold). During experiments, the pressures are typically at 1$\cdot$10$^{-5}$~mbar in the source chamber and at 4$\cdot$10$^{-8}$~mbar in the detection chamber.

A supersonic beam of \ce{H} atoms is generated by expanding a mixture of \ce{NH3} and \ce{Kr} (10~\% mixing ratio, 4~bar) through a pulsed valve into the vacuum. The \ce{NH3} is subsequently photolyzed using an ArF excimer laser (Neweks PSX-100) inside a quartz capillary (1~mm inner diameter, \mbox{15~mm} length) that is attached to the front of the valve. The 193~nm laser beam, typically at 2~mJ/pulse, is focused into the capillary close to the valve orifice using a cylindrical lens ($f$~=~350~mm).
The valve (General Valve Series 99, opening time of \mbox{$\approx$~230~$\mu$s} full width at half maximum) is mounted on a translation stage, thus allowing for adjustment in all three spatial dimensions. The valve body is cooled with a flow of gaseous nitrogen through a liquid-nitrogen reservoir, and its temperature is controlled through a feedback loop with a PID-driven air-process heater. In the deceleration experiments reported here, the valve temperature was set to 238~K and it remained stable to within 3~K.
The atoms pass through a skimmer that is placed at a distance of 50~mm from the nozzle. Thereafter, they are subject to the solenoid magnetic fields of the decelerator (first coil 85~mm from the valve). A detailed description of the Zeeman decelerator is given in Section \ref{sec:zeemandecelerator}.

After Zeeman deceleration, the \ce{H} atoms are photoionized in a (2+1) resonance-enhanced multiphoton ionization (REMPI) scheme via the 2s~$^2$S$_{1/2}$ state and the resulting ions are detected in a Wiley-McLaren-type TOF mass spectrometer. The extraction plates are turned on after the detection laser pulse to prevent losses from electric-field-induced mixing between the $^2$S$_{1/2}$ and the $^2$P$_{1/2}$ states. For the REMPI process, laser radiation at 243~nm is generated by frequency-doubling the output of a pulsed dye laser (Spectra Physics PDL3, Coumarin 480 dye) pumped by the third harmonic of a Nd:YAG laser (Quanta Ray, GCR 290) at 355~nm. The 1.2~mJ/pulse output energy is tightly focused into the chamber ($f$~=~150~mm) to increase the number of hydrogen ions produced. 

%Fig.~\ref{fig:wide}%
%\begin{figure*}
%\includegraphics{fig_1.eps}
%\caption{\label{fig:wide}Use the \texttt{figure*} environment to get a wide
%figure, spanning the page in \texttt{twocolumn} formatting.}
%\end{figure*}
%is too wide for a single column, so instead the \texttt{figure*} environment has been used.

\subsection{\label{sec:zeemandecelerator} Zeeman Decelerator Design and Properties}

The inhomogeneous magnetic fields for Zeeman deceleration are generated in a set of 12 solenoid coils. A maximum on-axis magnetic field of 2.5~T should be sufficient to entirely remove the initial kinetic energy of a Kr-seeded, supersonic beam of hydrogen atoms that exits the nozzle with a forward velocity of about 500~m/s (corresponding to 11~cm$^{-1} hc$ of kinetic energy). The electronic modules for the fast switching of the magnetic fields are based on the original circuit diagrams from the Merkt group \cite{Wiederkehr2011}. Both the mechanical parts and the electronics for the decelerator were built in-house in Oxford.

Concerning the mechanical construction, we opted for a modular design that allows us to both shape and to replace each coil individually, and hence to change quickly between different coil configurations, e.g., single coils with more turns, other radii or lengths. Due to the rather extreme demands on the coils from high-current pulsing, ease of exchangeability is also beneficial from an experimental point of view. Furthermore, every coil is connected to its own independent driver electronics, so that different sets of electronics can be used to operate the decelerator. For the focusing experiments presented in this article, we took advantage of this feature in particular.

Each coil is wound around a PEEK bobbin (6~mm inner diameter, 500~$\mu$m wall thickness) and cast into an aluminum shell using thermally conductive epoxy resin. Notches are cut into the aluminum housing to suppress eddy currents (see photograph in Figure~\ref{fig:sketchchamber}). The technical specifications of the decelerator coils (7~mm inner diameter, 70 turns in 4 layers, 400~$\mu$m diameter Kapton-insulated copper wire) are similar to those used by the Merkt group. Using an LHC bridge, we measured a resistance of $R$~=~0.29~$\Omega$ and an inductance of $L$~=~26~$\mu$H which is in accordance with calculations.

The coils are mounted onto a water-cooled aluminum base plate inside the vacuum chamber, as shown in the photograph in Figure~\ref{fig:sketchchamber}. A closed-cycle chiller system, operating with an ethylene glycol/water mixture to prevent corrosion of the base plate, keeps the coils at temperatures below 315~K even when pulsed at a 10~Hz repetition rate with currents of 300~A. Coil temperature and water flow through the decelerator are continuously monitored with an interlock system that automatically turns off the electronics in the event of a fault. The decelerator is mounted onto a supporting structure with screws for precise alignment in the $xy$ dimension. Base plates for additional deceleration stages can be added to the current setup to allow for the deceleration of particles with a smaller magnetic-moment-to-mass ratio than atomic hydrogen.

\begin{figure}
\includegraphics{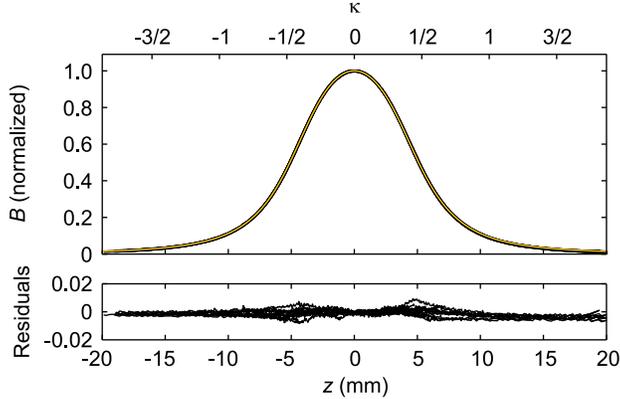}
\caption{\label{fig:staticBfield} Top: On-axis magnetic field of each solenoid coil in the experiment measured with a Hall probe sensor (black curves) at 6~A DC. The analytical solution is shown as a yellow curve. The data are normalized and centered to the maximum magnetic field of each coil along the beam axis \textit{z}. The abscissa on top of the graph gives a scale in terms of the reduced position $\kappa$ which is used to define the switch-off time of a coil for Zeeman deceleration (see Section \ref{sec:decelexperiments}). Bottom: Residuals between measured data and theory.}
\end{figure}

\begin{figure}
\includegraphics{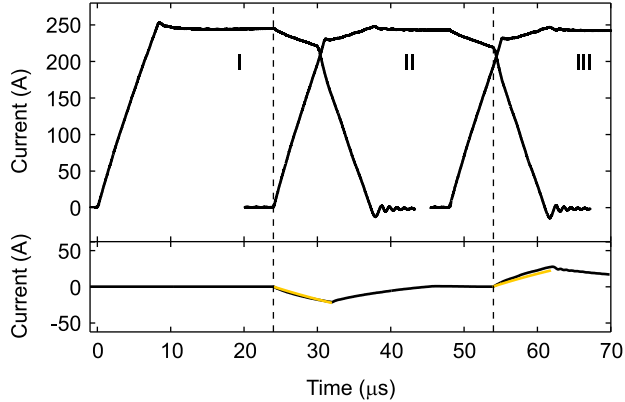}
\caption{\label{fig:mutualinductance} Temporal characteristics of the pulsed currents through coils I, II and III measured with a current probe. Coil I is the first coil after the skimmer. Due to mutual inductance effects, the current through a coil is decreased when it is temporally overlapped with an adjacent coil. The oscillation at the end of each pulse is an artifact of the current-probe measurement. \cite{Wiederkehr2011} The black curve in the lower part of the figure shows the current induced in coil I when coil II is pulsed (all other coils turned off). The induced current can be explained through RL circuit theory, as illustrated by the yellow curves.}
\end{figure}

We determined the real on-axis magnetic fields using a Hall probe sensor (Honeywell, SS59ET) that was moved through the decelerator assembly with a high-precision linear translation stage. To obtain magnetic field profiles, as shown in Figure~\ref{fig:staticBfield}, a static current of 6~A DC was applied to one coil at a time while the translation stage was shifted. The results are consistent with the nominal center-to-center coil distance $d$ of 10.7~mm (with a root-mean-square deviation of 90~$\mu$m). The experimental data for all 12 coils are accurately matched by the analytic solution \cite{Bergeman1987} to a 70-turn coil of 8.5~mm length, 7~mm inner diameter and 10.6~mm outer diameter. As illustrated by the residuals in Figure~\ref{fig:staticBfield}, the agreement between theory and experiment is so good that we can rule out any influence of the entering and exiting wires on the shape of the solenoid magnetic fields.

As described by Wiederkehr et al. \cite{Wiederkehr2011}, the fast switching of currents is achieved with an electronic circuitry based on insulated-gate bipolar transistors (IGBT). Briefly, a high-voltage power supply is used to obtain a rapid increase/decrease of the current (kick interval), while a constant current level is maintained via a low-voltage supply until the coil is switched off again (hold interval). The end of the kick phase and the current through a coil during the hold interval are determined by a comparator circuit which correlates the voltage drop through a shunt resistor with a preset value. In the experiments described here, we worked at a kick voltage of 800~V and a hold voltage of 110~V, giving rise to decelerator currents of $I$~=~243~A during the hold interval and rise and fall times of \mbox{8~$\mu$s} during the kick phase, as measured with a current probe (LeCroy CP500). % Measurements with a pick-up coil confirmed that the magnetic field closely follows the temporal behavior of the current.
Current pulses of adjacent coils were overlapped by 6~$\mu$s to prevent Majorana spin-flip transitions \cite{Hogan2007} during the deceleration process (see Section \ref{sec:results}). Under these conditions, we observe additional cusps in the current profile that are caused by mutual inductance effects (see Figure~\ref{fig:mutualinductance}). Using an LHC bridge, we measured a mutual inductance of $M$~=~2.8~$\mu$H which is in accordance with the output of finite-elements calculations (FEMM 4.2). The lower part of Figure~\ref{fig:mutualinductance} shows the current induced in coil I when coil II is switched; the temporal behavior of the induced current matches the switching times for coil II, as indicated by the dashed vertical lines. The induced current in coil I can be approximated by RL-circuit theory, thus yielding an exponential growth (yellow curves in Figure~\ref{fig:mutualinductance}) with a characteristic time $L/R$ and an amplitude ${M I}/{R t_\textnormal{r}}$, where $t_\textnormal{r}$ is the rise time.

Mutual inductance effects are intrinsic to every setup that consists of coils in close proximity whose currents are rapidly changed. The influence of mutual inductance can be reduced by increasing the coil distance, e.g., $M\approx$~0.4~$\mu$H at a center-to-center coil distance of 2~$d$. However, this would require longer switching times for the coils to prevent losses from Majorana transitions and a redistribution in phase space due to free flight, and it would also come at the expense of a much longer apparatus. Nonetheless, mutual inductance effects do not pose a major limitation on Zeeman deceleration. The cusps in the current profiles can be taken into account both in the generation of the pulse sequence for Zeeman deceleration and in the three-dimensional particle trajectory simulations. For a 6~$\mu$s pulse overlap, the current through a coil is about 10~\% higher than its preset value, because it is still rising when the comparator circuit switches from the kick to the hold interval. The change in current can be accounted for in the generation of the pulse sequences for Zeeman deceleration. Likewise, it is possible to adjust the reference voltage for each comparator to the output of a current probe, such that the current during the hold interval matches the current of the other coils. 
% with a potentiometer

\subsection{\label{sec:focexpexp} Transverse Focusing Experiments}

In the experiments described here, eleven coils were operated with the decelerator electronics. One coil was disconnected from the decelerator circuitry and driven by another set of electronics that provided a comparably low current for an extended period of time (``quasi-static'' operation). In the following, this coil will be referred to as the ``focusing coil''.
For these electronics, the temporal current profile is described by an RL circuit with $R$~=~\mbox{0.95~$\Omega$} for experiments with pulse durations of 600~$\mu$s, and $R$~=~\mbox{3.15~$\Omega$} for measurements in which the delay of the focusing coil was scanned (pulse duration of 100~$\mu$s). The current rises/decays exponentially, with a characteristic time $L/R$ and a current amplitude $I_{\textnormal{max}}\leq60$~A. The higher resistance in the latter set of experiments was used to decrease the length of the trailing edges of the current pulse.

Most of the data were taken with coil VI for focusing, but coils I and XII were also used for comparison in a number of additional experiments (labeling the coils from I to XII in the order in which the beam passes through them). Typically, the current direction was the same for all coils. However, for some measurements, we also reversed the direction of the quasi-static current as denoted with a negative sign for the current of this coil. % By convention, we use a positive sign for the applied current if its direction is the same as for the deceleration coils and a negative sign for the reverse current direction.

All data were taken at a 10~Hz repetition rate and time-of-flight traces with and without magnetic fields were compared by alternating the applied field on and off on a shot-by-shot basis. For experiments with the focusing coil, a three-shot sequence was used to get a reference measurement in which current was only applied to the deceleration coils. The reference was recorded using the same time sequence for Zeeman deceleration to avoid changes in the TOF profile due to secondary effects caused by a change in decelerator timings.

In general, the data are very reproducible and have little noise. A complete TOF scan, as shown for example in Figure~\ref{fig:coil6}(a), was usually completed within 20 minutes. Each data point typically corresponds to an average of 50 laser shots.

\section{\label{sec:results} Results and Discussion}

\subsection{\label{sec:decelexperiments} Interpretation of TOF Data}

The operation of a Zeeman decelerator is based on phase stability, and the principles are similar to other supersonic beam deceleration techniques. Detailed descriptions of phase stability are, for example, given by Bethlem et al. \cite{Bethlem2000} for Stark deceleration, and by Hogan et al. \cite{Hogan2007} and Wiederkehr et al. \cite{Wiederkehr2010} for Zeeman deceleration. In this article, we give a brief summary of the basic ideas and point out differences with respect to previous definitions.

As mentioned in the Introduction, Zeeman deceleration exploits the interaction of paramagnetic atoms and molecules with time-varying, inhomogeneous magnetic fields. By successively switching high currents inside an array of solenoid coils, kinetic energy is converted into Zeeman energy, and permanently removed when a coil is switched off.
The coil switching times are chosen such that a so-called ``synchronous particle'' has always moved by exactly one coil distance $d$ in the time interval between one coil and the next being switched off.
A range of other (non-synchronous) particles, which are further ahead or behind the synchronous particle, are decelerated as well. Faster particles move further into the solenoid magnetic field, and are therefore more strongly decelerated, while slower particles are decelerated less. Non-synchronous particles will oscillate about the position and velocity of the synchronous particle, provided that they are within the phase-stable region of the decelerator.

By analogy with Stark deceleration and particle acceleration schemes, we use a reduced position $\kappa=z/d$ to describe the position of the synchronous particle relative to the center of a coil (see Figure \ref{fig:staticBfield}). The reduced position of the synchronous particle after the switch-off, i.e., when the decreasing current reaches zero, is denoted as $\kappa_0$. In previous studies on Stark and Zeeman deceleration, a phase angle $\phi_0$ has been used to describe the operation of the decelerator, as in, e.g., Wiederkehr et al. \cite{Wiederkehr2010}. However, this notation is rather misleading in terms of Zeeman deceleration, because the shape of the on-axis potentials is not periodic thus resulting in non-zero deceleration at $\phi_0$ = 0 and counter-intuitive phase angles ($<$~0 or $>$~2~$\pi$). The conversion between $\kappa_0$ and $\phi_0$ is given by $\kappa_0 = \phi_0/\pi + v t_\textnormal{r}/d - 1/2 $. Due to the non-zero ramp time $t_\textnormal{r}$, there is an explicit dependency on the velocity $v$ of the synchronous particle. Hence, deceleration at constant $\kappa_0$ is not equivalent to operation at a constant phase angle $\phi_0$. 

The required pulse sequence to be applied to the decelerator is calculated in a one-dimensional numerical trajectory simulation using the positions and velocities of the synchronous particle. In our case, this synchronous particle is a ground-state hydrogen atom in the low-field-seeking $M_F$~=~1 Zeeman sublevel, where $M_F$ denotes the projection of the total angular momentum $F$ onto the local magnetic field axis.
The influence of the magnetic fields on the particles is monitored by their time of flight (TOF) through the apparatus. In our experiments, the time of flight of the hydrogen atoms is defined by the relative temporal delay between the excimer laser pulse for photolysis and the pulsed UV laser beam for photoionization.

The TOF data are interpreted using numerical three-dimensional particle trajectory simulations. In our code, written in MATLAB and based on similar concepts to other programs \cite{Salathe2009}, the geometry, timings and magnetic fields are matched as closely as possible to those in the experiment. A Monte-Carlo approach is used to select the initial positions and velocities of four million trajectories, with the particles
equally distributed amongst the four Zeeman states (two high-field seeking and two low-field seeking) of \ce{H} 1s $^2$S$_{1/2}$. % An initial selection step ensures that only those particle trajectories that will pass through the skimmer are selected.

The hydrogen atom beam can be modeled as a supersonic jet with a Gaussian velocity distribution centered at an initial longitudinal velocity of 490~m/s. The beam velocity was derived by comparing peak positions of simulated and experimental TOF profiles for Zeeman deceleration at different switching times. Note that the mean velocity of the beam can be different from the selected initial velocity 
of the synchronous particle, $v_0$, that is used for the generation of the pulse sequence; for example, a below-average initial velocity may be selected to achieve a lower final velocity after deceleration. For the experiments described in this article, we typically use $v_0$~=~500~m/s. The transverse velocity distribution of the beam is modeled as a bivariate normal distribution centered at zero. The 
widths of the distributions are inferred from the longitudinal beam temperature, $T_{z}$, and the transverse beam temperature, $T_{r}$; neither of these temperatures were measured experimentally. However, the width of the TOF profiles gives a good estimate for $T_{z}$ and, owing to the selectivity of the skimmer, the simulated TOF profiles are not markedly sensitive to $T_{r}$, even when the transverse temperature at the source is increased from 10~mK to 150~mK. In the simulations for this article, we use $T_{z}$~=~\mbox{1.1~K} and $T_{r}$~=~\mbox{10~mK}. 

At a source temperature of 238~K, the initial beam velocity is about 150~m/s higher than the velocity of a pure supersonic jet of krypton atoms \cite{Christen2008}. This indicates that, probably due to the large mismatch in mass between H and the Kr carrier gas, the excess translational energy of the H atoms after 193~nm laser photolysis ($\approx$~4500~cm$^{-1} hc$ on average \cite{Koplitz1987}) does not fully thermalize through collisions during the supersonic expansion. Related to that, we found that the experimental TOF data could only be matched in simulations by assuming that the initial particle positions were uniformly distributed over a region equal to the length of the capillary, even though the photolysis laser beam is only 2.5~mm in diameter. The evolution of the supersonic expansion is also seen in a time delay after photolysis, which was experimentally determined as 31~$\mu$s, before the atoms emerge from the capillary.

The pulsed magnetic fields of the decelerator are modeled using analytical approximations to the observed current profiles and static magnetic fields (Figures~\ref{fig:staticBfield} and \ref{fig:mutualinductance}). Particle positions and velocities are integrated at a numerical time step of 100~ns using the Velocity Verlet algorithm \cite{Swope1982}. Simulations at a 10~ns time step give very similar results, but come at the expense of computational time.

\begin{figure}
\includegraphics{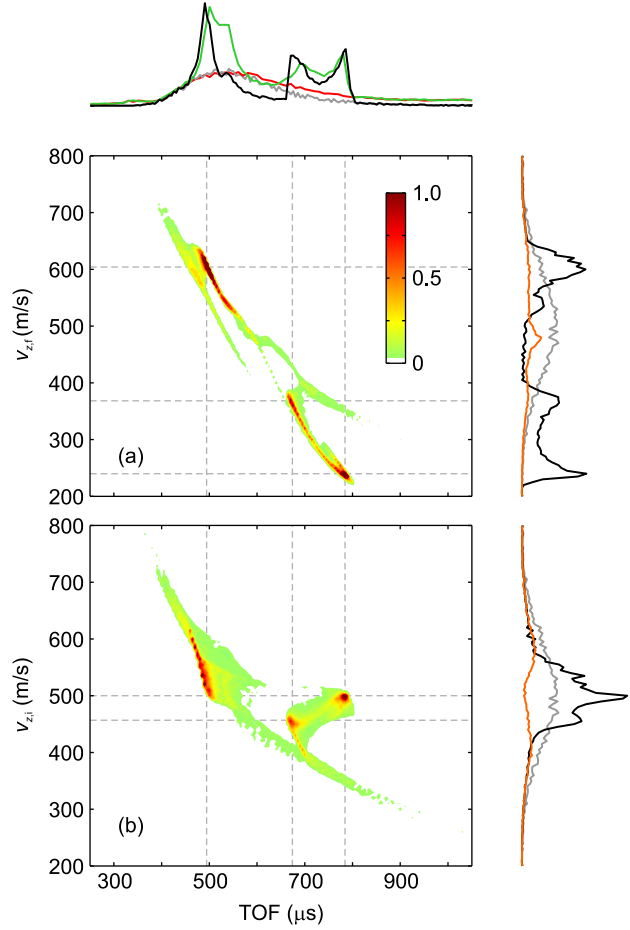}
\caption{\label{fig:TOFexpl} Zeeman deceleration from 500~m/s to 240~m/s at $\kappa_0$~=~0. The focusing coil (coil VI) is not switched. Top: Experimental (green trace) and simulated (black trace) TOF profiles. Measured and calculated TOF traces without the influence of the decelerator magnetic fields are shown as red and grey curves, respectively. Density plot (a): simulated longitudinal particle velocities after deceleration, $v_{z,\textnormal{f}}$, versus arrival time. Density plot (b): simulated initial longitudinal particle velocities of those trajectories arriving at a given TOF, $v_{z,\textnormal{i}}$, versus TOF. Density scale for (a) and (b) in arbitrary units. Right: Distribution of longitudinal particle velocities before (bottom right) and after (top right) Zeeman deceleration subdivided into particles in low-field-seeking (black curves) and high-field-seeking states (orange curves). The particle velocity distributions without magnetic fields are given as gray curves for comparison. All plots are scaled with respect to the maximum TOF signal without deceleration (gray curve at the top). Note that only the velocities and arrival times of particles in the focus of the detection laser are taken into account.}
\end{figure}

In Figure~\ref{fig:TOFexpl}, the output of a three-dimensional numerical trajectory simulation for Zeeman deceleration from 500~m/s to 240~m/s at $\kappa_0$~=~0 is shown. The time-of-flight traces for deceleration are compared with the experimental data at the top of the figure, and have a rather complicated structure which is due to the complex particle motion inside the decelerator. % Neither the initial nor the final longitudinal velocities of {\it detected} particles form a Maxwell-Boltzmann distribution (black and orange curves on the right), as would be the case without a magnetic field present (grey curves on the right). 
The plots in (a) and (b) show the distributions of final and initial longitudinal velocities that lead to the time-of-flight signal above, respectively. As shown in (a), the peak at around 500~$\mu$s is due mainly to particles that are accelerated from 500~m/s to 600~m/s; the TOF signal at 680~$\mu$s originates from particles that are partly decelerated from 460~m/s to 370~m/s. Only the peak at an arrival time of 780~$\mu$s can be attributed to the particles fully decelerated to 240~m/s, with a minor contribution of particles traveling at a velocity of 370 ~m/s. 
The velocity distribution of the decelerated atoms is very narrow, with a half-width of approximately 20~m/s.
As shown in (b), the fully decelerated particles originate almost exclusively from a velocity distribution of low-field-seeking particles centered at the selected velocity for the synchronous particle (500~m/s). 

At the right-hand side of panel (a), the velocity distributions of particles in low-field-seeking quantum states and high-field-seeking quantum states are plotted as black and orange curves, respectively. In the experiment, we cannot spectroscopically distinguish between the different quantum states, but the results from trajectory simulations strongly suggest that the number of transmitted high-field seekers is very small, confirming the good quantum-state selectivity of the deceleration process. Particles in high-field-seeking states are subject to forces opposite to those experienced by low-field seekers. Hence, their motion is not phase stable, and they are deflected towards the even-higher off-axis magnetic fields.
% An experimental TOF profile (green curve) for Zeeman deceleration is superimposed with the output of the trajectory calculation (black curve) at the top of Figure~\ref{fig:TOFexpl}. For comparison, the red and grey traces correspond to measured and simulated traces in which no magnetic fields are switched.

The agreement between theoretical and experimental TOF traces is very good. However, a higher signal level is measured in between the main peaks (from about 550~$\mu$s to 650~$\mu$s) than predicted by the simulation. This increase in signal could be a sign of population redistribution among the Zeeman sublevels, induced by Majorana spin-flip transitions \cite{Hogan2007}. Majorana losses are caused by a rapid reversal of the magnetic field direction, and may be driven by a current undershoot during the kick interval at switch-off \cite{Wiederkehr2011}. This argument is further supported by trajectory simulations in which we included state-to-state transition probabilities as a rough estimate of the contribution from Majorana spin-flip transitions during deceleration (not shown). The transition probabilities were derived from a numerical solution to the time-dependent Schr\"{o}dinger equation for ground-state atomic hydrogen (including hyperfine structure) as a function of slew rate through a zero magnetic field. Using this spin-flip approximation in the simulation, we observed a considerable signal increase in between TOF peaks. However, the overall fit to the TOF profiles was not improved, probably due to the inherent approximations in the spin-flip modeling and its implementation in the trajectory code. In any case, as detailed in Section IIIB, Majorana transitions do not appear to have a significant influence on the experimental results on transverse focusing that are presented in this article. % Therefore, simulation results shown in this article do not take Majorana spin-flip transitions into account.

The final velocity of the hydrogen atoms after Zeeman deceleration can be tuned by changing $\kappa_0$, as illustrated in Figure \ref{fig:coil6}. As $\kappa_0$ is increased from -1 to 0, the decelerated peak is shifted to later arrival times which also corresponds to a decrease in final particle velocity from 490~m/s at $\kappa_0$~=~-1 to 240~m/s at $\kappa_0$~=~0. At large positive $\kappa_0$, it is also possible to increase the particle velocity, e.g., at $\kappa_0$~=~2. In this case, an accelerating force is exerted onto the particles as they pass through the decreasing magnetic field of a coil (Figure \ref{fig:staticBfield}).

The comparison between experiment and simulation in Figure \ref{fig:coil6} generally shows good agreement in terms of the positions, widths and relative intensities of peaks in the time-of-flight spectrum.

\subsection{\label{sec:focexp} Transverse Focusing Experiments}

\begin{figure}
\includegraphics{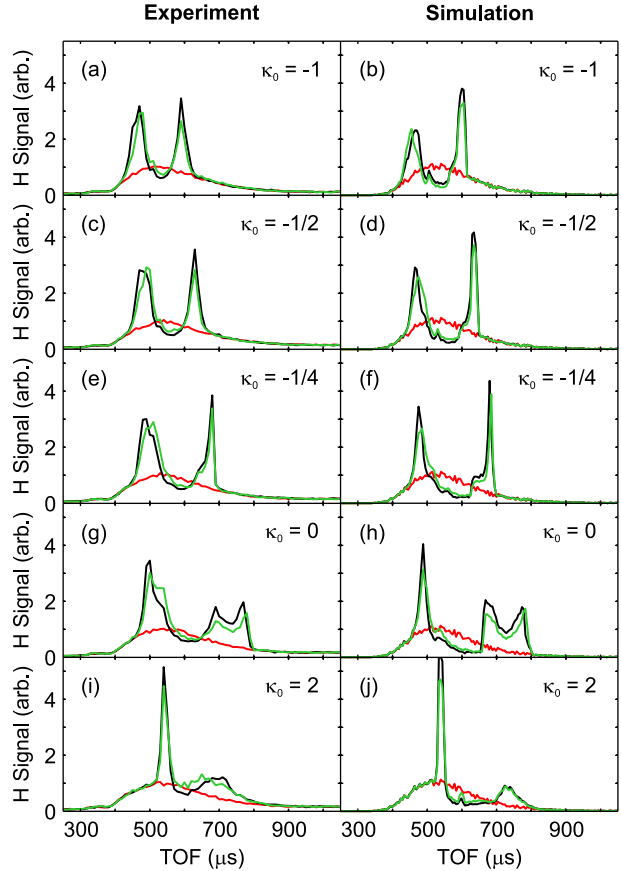}
\caption{\label{fig:coil6} Experimental (left column) and simulated (right column) TOF profiles for Zeeman deceleration/acceleration from 500~m/s to a range of different final velocities of the synchronous particle. In the green traces, only the deceleration coils are switched. For the TOF profiles in black color, an additional current of -30~A is applied to the focusing coil (coil VI). The red curves correspond to traces in which no magnetic fields are present. The final velocities are: (a) and (b) 490~m/s ($\kappa_0$~=~-1), (c) and (d) 440~m/s ($\kappa_0$~=~-1/2), (e) and (f) 350~m/s ($\kappa_0$~=~-1/4), (g) and (h) 240~m/s ($\kappa_0$~=~0), (i) and (j) 590~m/s ($\kappa_0$~=~2).}
\end{figure}

\begin{figure}
\includegraphics{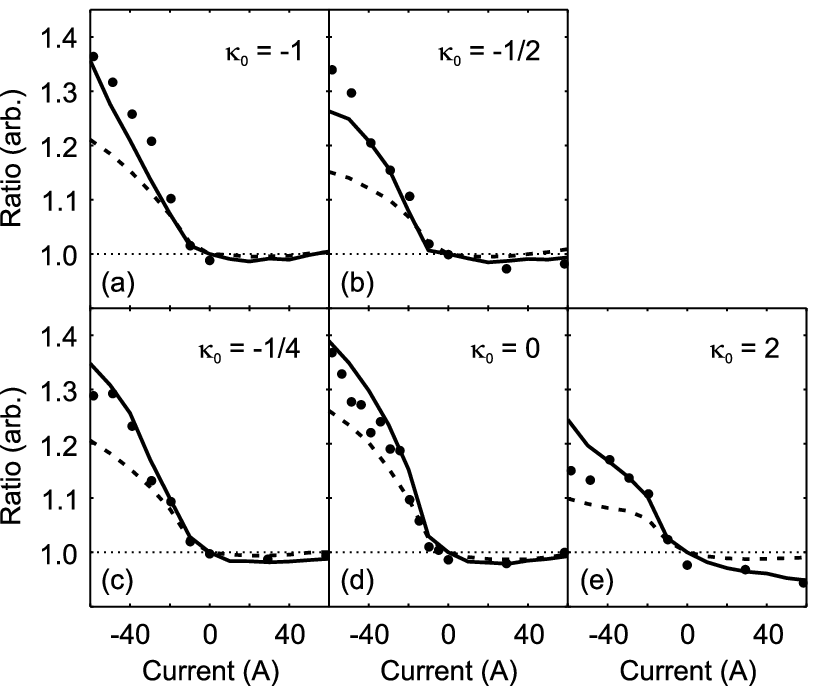}
\caption{\label{fig:biascurrent} Change in decelerated hydrogen atom signal as a function of current applied to the focusing coil (coil VI). The ratio is derived from a comparison between the integrated signal with and without current through the focusing coil. Experimental data are shown as dots. Results from trajectory simulations are shown as solid and dashed lines. For the dashed lines, the ratio is obtained by counting all particles that pass through an imaginary detection area spanned by the $xy$ plane perpendicular to the beam axis. For the solid lines, only particles in the focus of the laser beam (a cylinder of 0.5~mm diameter assumed) are taked into account.}
\end{figure}

In this section, we demonstrate that the transverse particle confinement in a Zeeman decelerator can be improved by applying a negative, quasi-static current to a coil that would otherwise be used for deceleration. This effect was studied using various experimental settings. In the following, we analyze the dependency of the signal enhancement on $\kappa_0$ and the influence of the magnitude and the direction of the applied quasi-static current. By varying the position and the switch-on time of the focusing coil, we show that the observed effects are indeed related to the action of this specific coil only.

As illustrated in Figure \ref{fig:coil6}, there is an increase in signal intensity for all peaks in the TOF trace -- typically 10--30~\% of peak height -- when a quasi-static current of -30~A is applied to coil VI (black traces). This effect is observed for all values of $\kappa_0$, and it is well reproduced by particle trajectory simulations. The signal increase becomes even more apparent in Figure \ref{fig:biascurrent}, where currents of up to $\pm$60~A were used. For the largest negative values of the current, the integrated signal enhancement is 30--40~\% for all $\kappa_0$ values in the range -1 to 0.
In this figure, a signal ratio is plotted as a function of current applied to the focusing coil. The ratio is derived by comparing, on a shot-by-shot basis, the integrated signal within a specified time gate encompassing the decelerated peak in the TOF (see Figure \ref{fig:coil6}) with and without current applied to coil VI. We checked the outcome using different widths and positions of the arrival-time gates for integration, and we obtained results that are very similar to those shown in Figure \ref{fig:biascurrent}. 
We also carried out measurements at zero current for the focusing coil, in which case we would expect a ratio of unity as the current is the same for both measurements. The observed deviation from 1.0 therefore gives a good estimate of the statistical error in the experiment.

Figure \ref{fig:biascurrent} illustrates that the spatial extent of the detection laser beam needs to be taken into account in the data analysis (solid lines) in order to see good agreement between experiment and simulation. The ratios obtained by taking into account all decelerated particles passing beyond the final stage of the decelerator (dashed lines) are too small to explain the signal gain observed in the experiment.
%The observed boost in signal is not merely an increased focusing of particles into the focal point of the detection laser beam. The dashed lines in Figure \ref{fig:biascurrent} show that there is an increase in the overall number of decelerated particles transmitted through the decelerator, typically about 15--25~\% at -60~A. When taking the detection (laser-focus) volume into account, the gain on the beam axis accounts for another 10--15~\% such that a total signal increase of up to 40~\% is observed (solid lines).
Thus, the observed boost in signal has two origins: an increase in the overall number of decelerated particles transmitted through the decelerator, typically about 15--25~\% at -60~A; and a stronger focusing of particles into the detection laser volume, which accounts for another 10--15~\% increase in signal.

\begin{figure}
\includegraphics{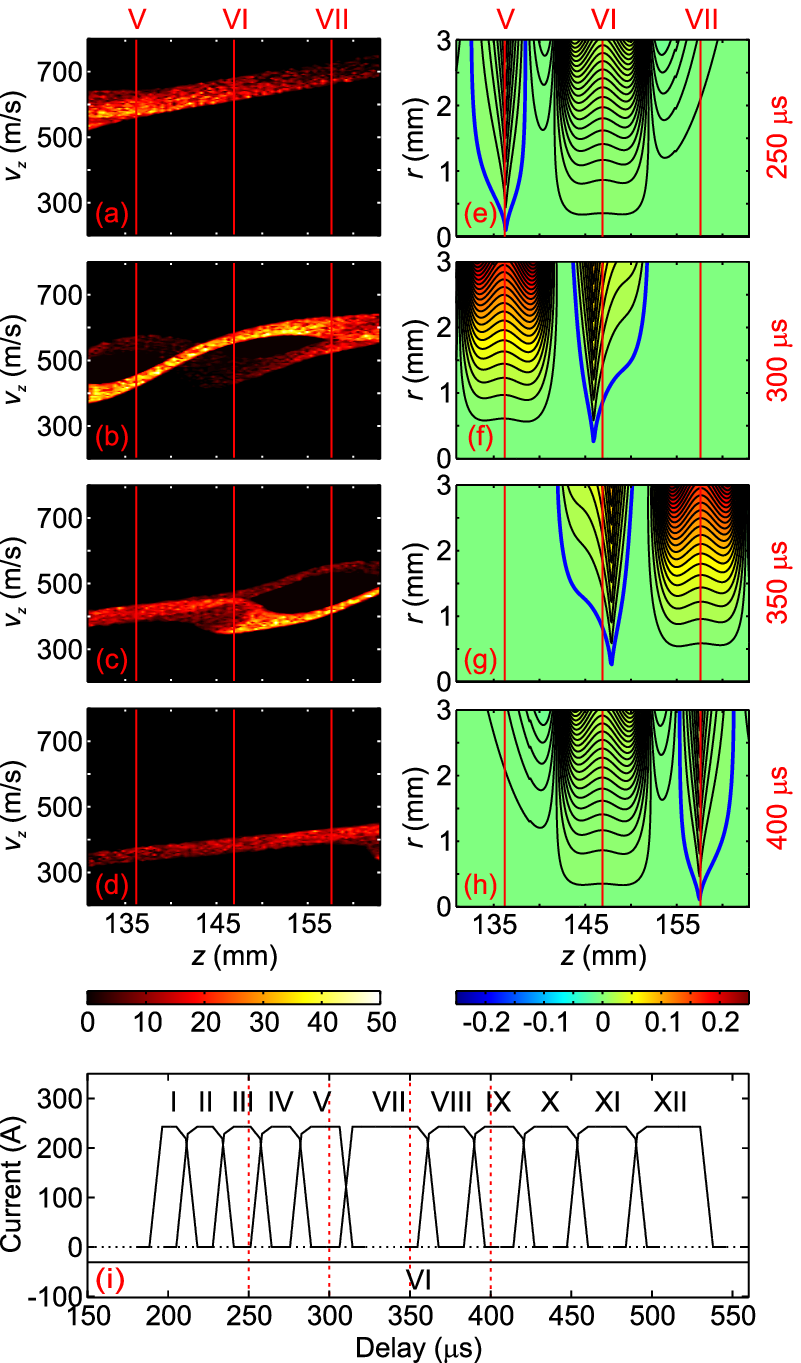}
\caption{\label{fig:biasdelaysim} Left column (a)--(d): Snapshots of the simulated longitudinal particle velocities $v_{z}$ along the beam axis $z$ for Zeeman deceleration at $\kappa_0$~=~0 using coil VI for transverse focusing. The snapshots are taken (a) 250~$\mu$s (during pulse III), (b) 300~$\mu$s (pulse V), (c) 350~$\mu$s (pulse VII) and (d) 400~$\mu$s (pulse IX) after the photolysis laser shot (times marked with dashed vertical lines in (i)). Only particles arriving in the detection region are plotted. Right column (e)--(h): Contour plots of the relative magnetic field $\Delta B(r,z)=B(r,z)-B(r=0\textnormal{~mm},z)$ (in T), where $r$ is the transverse distance from the beam axis, given at the same times as in (a)--(d). The center positions of the focusing coil and its neighboring coils are indicated with red vertical lines. The extent of the focusing field is shown with blue contour lines. (i): Schematic current pulse sequences for the deceleration coils and the focusing coil. A current of -30~A is applied to coil VI for the entire duration of the deceleration pulse sequence (600~$\mu$s).}
\end{figure}

In general, the signal ratio increases as the magnitude of the negative current through coil VI is raised. On the other hand, there is no increase in signal if the current direction for the focusing coil is the same as for the deceleration coils. The observed trends can be explained by a change in the shape of the transverse magnetic field between coil VI and the adjacent deceleration coils. The anti-parallel currents generate quadrupole-like magnetic fields between coils V and VI and between coils VI and VII (see right column in Figure \ref{fig:biasdelaysim}), which are more strongly focusing in the transverse beam direction than in the case of two deceleration coils with parallel currents. 
The quadrupole-like field is not generated if a positive current is applied to coil VI, hence there is no additional transverse confinement and no signal enhancement.

The complex interplay between the magnetic fields and the motion of the particles in the Zeeman decelerator makes it difficult to find simple physical explanations for differences in the signal ratios for the different values of $\kappa_0$ shown in Figure \ref{fig:biascurrent} (a)--(e). 
We observe much smaller ratios for acceleration ($\kappa_0$~=~2), even though the average longitudinal and transverse confinement is comparable to other reduced positions, e.g., to deceleration at $\kappa_0$~=~-1/2. The differences might be related to the different mode of operation (coil switched off when the synchronous particle is moving down the potential hill) which makes the transverse focusing with coil VI more effective for deceleration than for acceleration.

\begin{figure}
\includegraphics{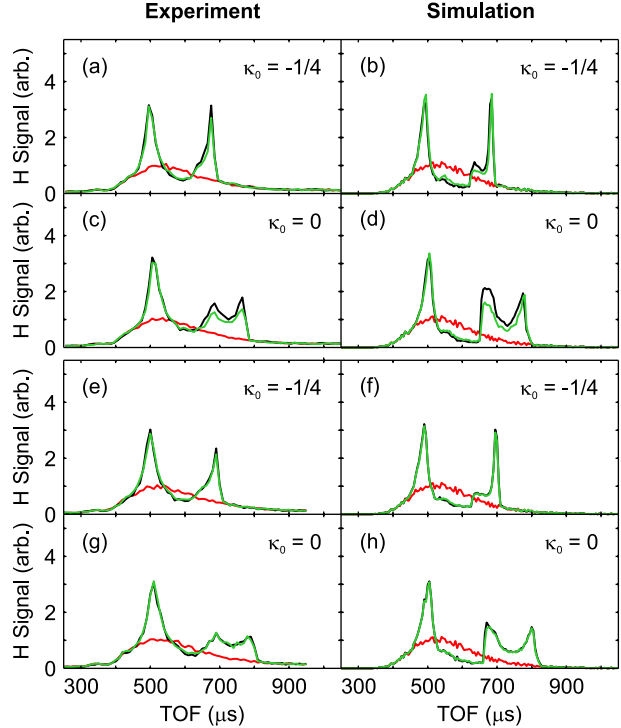}
\caption{\label{fig:coil1} Experimental (left column) and simulated (right column) TOF profiles for Zeeman deceleration from 500~m/s to 350~m/s ($\kappa_0$~=~-1/4) and 240~m/s ($\kappa_0$~=~0) using coil I in (a)--(d) and coil XII in (e)--(h) at a current of -30~A to study transverse focusing effects. The color code is the same as in Figure \ref{fig:coil6}.}
\end{figure}

We also studied changes in the time-of-flight profiles when coils I and XII, rather than VI, were used for transverse focusing. At a focusing current of -30~A, we observe an increase in the decelerated atom signal at $\kappa_0$~=~-1/4 and 0 when coil I is used for focusing (Figure \ref{fig:coil1}). However, in contrast to the experiments with coil VI, there is no change in the signal intensity of the peak at 500~$\mu$s that can be assigned to accelerated hydrogen atoms. For coil XII, we do not see any noticeable change in the TOF traces. These effects can be understood in terms of the quadrupole magnetic fields in the transverse direction that are generated during the switching of a deceleration coil and the simultaneous operation of the focusing coil in a quasi-static mode, as illustrated in Figure \ref{fig:biasdelaysim} for focusing with coil VI. % The left column in this figure depicts snapshots of the simulated longitudinal phase-space distribution at four different times after photolysis. Plots of the relative magnetic field with respect to the beam axis, $\Delta B(r,z)=B(r,z)-B(r=0\textnormal{~mm},z)$, are shown in the right column. Vertical dashed lines in Figure \ref{fig:biasdelaysim} (i) visualize which coils are turned on at each specific time. 
As the deceleration coils are successively pulsed, the position of the trapping minimum (enclosed by the blue contour lines) is shifted along the beam axis, in Figure \ref{fig:biasdelaysim} (e)$\rightarrow$(h) from $z$~=~136~mm to 158~mm. If current is applied to both coils V (or VII) and VI, the difference between the on-axis and off-axis magnetic fields, $\Delta B(r,z)$, is equal to 0.1~T (at $r$~=~3~mm). For hydrogen atoms in low-field-seeking states, this magnetic field is already sufficient to confine particles with a transverse velocity of 35~m/s. Figure \ref{fig:biasdelaysim} also illustrates that the generation of transverse focusing fields is not restricted to neighboring coils only. 
A small transverse magnetic trapping field of about 0.01~T -- sufficiently high to confine hydrogen atoms with off-axis velocities of up to 11~m/s -- is created even when the position of the focusing coil is further away from a deceleration coil, e.g., for coil III (as in (e)) or coil IX (as in (h)).

The additional transverse focusing fields at $z$~=~136~mm  in Figure \ref{fig:biasdelaysim} (f) and 158~mm in \ref{fig:biasdelaysim}(g) stem from the switching of deceleration coils V and VII, respectively. The magnetic field in the center of these coils is not seen by the decelerated particles, because each deceleration coil is turned off before the particles reach the coil center in order to maintain longitudinal phase stability. % prevent defocusing in the longitudinal direction. 
However, the transverse magnetic fields in conjunction with coil VI are still present when the slowed particles pass through this region in the Zeeman decelerator.

As can be seen from Figure \ref{fig:biasdelaysim} (a), the magnetic focusing field generated by simultaneously switching focusing coil $n$ and a deceleration coil $n-j$, where $j~\geq$~1, has an effect on the fast-moving atoms only. At this time in the deceleration sequence, the slow atoms will not have reached the position of the quadrupole yet. Likewise, only the slow atoms will benefit from the quadrupole field between the focusing coil $n$ and coils $n+j$ (Figure \ref{fig:biasdelaysim} (b)--(d)), since the fast atoms are already too far ahead in the Zeeman decelerator to be affected. In the case of coil I, there are no coils $n-j$; hence, when using I as the focusing coil, only the slow atoms can experience a focusing effect, and only an increase of the decelerated atom signal is observed. Likewise, there are no coils beyond coil XII, so that there is no transverse confinement for the slow atoms when this coil is used for focusing. In addition to that, the off-axis particles are already lost to the decelerator walls when reaching coil XII, so that the application of a focusing current to this coil has no effect at all.

Measurements in which the timing of the focusing coil is scanned relative to the start of the deceleration pulse sequence (not shown) further support these arguments. We observe a signal increase at earlier delay times for coil I than for coil VI, which is in accordance with the particle motion and the position of the coils inside the decelerator. Since coil I is at the front and coil VI is in the center of the Zeeman decelerator, it makes sense to switch coil VI at later delays to obtain a transverse focusing effect for the particles. We also see that the fast atoms are focused at earlier delays than the decelerated atoms, since they are further ahead in the decelerator, so that they experience the focusing fields at much earlier times.

Data at zero current for the focusing coil (Figure \ref{fig:coil1}) show that both free flight through the focusing coil and Majorana spin-flip transitions do not have an impact on the experimental results. In this case, the focusing coil is not used for deceleration, so that there is an additional region of free flight for this coil. If free flight had any effect on the experimental results, the TOF profiles with coil VI held at zero current would show less signal than applying zero current to coil I or coil XII, because free flight in between two deceleration stages is more detrimental to the longitudinal phase space than at the beginning or at the end of the deceleration pulse sequence. The TOF traces in Figure \ref{fig:coil6} (coil VI) and Figure \ref{fig:coil1} (coils I and XII) look very similar when no current is applied for transverse confinement (green curves). Hence, we can rule out that the observed increase in signal intensity for a negative current to the focusing coil is merely due to a compensation of particle loss from free flight. 

The potential influence of Majorana transitions on the shape of a TOF profile was described in Section \ref{sec:decelexperiments}. The use of a negative current for the focusing coils leads to the formation of quadrupole-like magnetic traps with a point of zero magnetic field along the beam axis. The resulting low magnetic field region does not substantially induce spin-flip transitions in our experiment. If Majorana transitions had a primary influence on these transverse focusing results, we would only be able to see these effects in the time-of-flight traces for coil VI. Assuming an equal distribution of quantum states after photolysis, spin flips in coil I would not cause a change in the quantum-state distribution. Likewise, coil XII is too close to the detection region to obtain a clear spatial separation of the high-field-seeking states that would cause a change in the TOF profile. The signal intensity of the decelerated atom peaks at 680~$\mu$s and 780~$\mu$s remains unchanged at zero focusing current confirming that Majorana losses are negligible.  In any case, the good fit shown in Fig. \ref{fig:biascurrent} between experiment and simulations that do not include Majorana transitions supports this argument.

\section{\label{sec:conclusions} Conclusions}

Over the past decades, magnetic quadrupole fields have been commonly used to trap paramagnetic atoms, particularly in conjunction with laser cooling \cite{Phillips1998}. Quadrupole fields have also found applications in the area of Zeeman deceleration. The magnetic trapping of hydrogen and deuterium atoms after Zeeman deceleration was demonstrated by pulsing two solenoid coils in an anti-Helmholtz configuration at a current of 200~A \cite{Hogan2008b, Wiederkehr2010a}. More recently, co-moving-trap Zeeman decelerators were developed in which the transverse confinement is either achieved using anti-Helmholtz coils \cite{LavertOfir2011, LavertOfir2011a} or a magnetic quadrupole guide \cite{Trimeche2011}. In comparison to the focusing experiments presented here, moving trap decelerators provide strong transverse confinement during the entire deceleration process. In the design from the Narevicius group \cite{LavertOfir2011, LavertOfir2011a}, transverse magnetic fields between 0.4 T and 1.2 T (at $r$~=~5 mm) were generated using peak currents of up to 500 A. The quadrupole guide used by Trimeche et al. \cite{Trimeche2011} was operated at 130 A, giving rise to a magnetic field of about 180 mT (at $r$~=~0.6 mm). Moving-trap decelerators are advantageous for trapping experiments after deceleration, because virtually all particles with a transverse velocity component are captured. However, a large transverse velocity spread is usually not desired in collision experiments or spectroscopic applications.

In this article, we have shown that the transverse acceptance in a Zeeman decelerator can be increased by applying a low current ($\leq$~60~A) to one of twelve deceleration coils, provided that the current direction is opposite to the other coils. With only a minor change in the coil configuration, we have already achieved a 40~\% increase in the signal intensity on the beam axis. At the same time, we have shown that the transverse focusing is much less effective when two adjacent coils carry currents in the same direction. % Intuitively, experimentalists may consider using strong quadrupole magnetic fields to achieve transverse focusing in their decelerator. In doing so, these fields have to be distributed over the entire length of the experiment to prevent losses from overfocusing. 
This implies that optimum transverse focusing can be attained without having to generate strong transverse focusing fields on the entire length of the decelerator. Our results suggest an alternative route towards attaining optimum transverse confinement, e.g., by inverting the current direction through every {\it n}th coil and pulsing it with a low, quasi-static current as described in this article. Similarly, the use of inverted current pulses may be considered. % On the one hand, the deceleration schemes must be carefully evaluated in terms of overall phase-space acceptance, and effects from Majorana transitions need to be considered as well. On the other hand, different coil configurations and pulse sequences can be implemented without much additional experimental effort. 

On the basis of simulations (to be reported in a future publication), we are convinced that more sophisticated focusing schemes will improve the transmission through the decelerator even further, and they can be used to tailor the height of the potential well in the transverse direction thus allowing for both trapping experiments and collision studies with the same device. % The overall gain with respect to the normal deceleration mode will depend on the desired final velocity. Wiederkehr et al. \cite{Wiederkehr2010} showed that their Zeeman decelerator was less efficient in removing little kinetic energy (phase angles $<$~30$^{\circ}$) over the entire length of the apparatus. Preliminary simulation results for an improved focusing scheme involving the reswitching of deceleration coils in the reversed current direction suggest a signal increase of about a factor of three; a factor of two is within reach at higher $\kappa_0$. A detailed discussion of this approach and the simulations would go beyond the scope of this article.

Our results do not only indicate an increase in the overall number of particles at the end of the decelerator, but also a more effective focusing towards the beam axis. This may become useful in experiments in which beam overlap is of major importance, for example, in crossed-beam machines or in cold collision studies between decelerated molecules and laser-cooled ions which are spatially confined in a sub-millimeter volume inside an ion trap \cite{Bell2009a}.

\begin{acknowledgments}

We thank the PTCL workshops in Oxford, especially Neville Baker and Andrew Green, for their commitment to this project. At ETH Z\"{u}rich, we are indebted to Hansj\"{u}rg Schmutz and Alex Wiederkehr for numerous fruitful discussions, and we thank Fr\'{e}d\'{e}ric Merkt for his efforts in taking this collaboration forward. K.D. acknowledges financial support from the Fund of the German Chemical Industry through a Kekul\'{e} Mobility Fellowship. M.M. is grateful for support through the ETH Fellowship Program. This work is financed by the EPSRC under Projects EP/G00224X/1 and EP/I029109/1.

\end{acknowledgments}

% \nocite{*}
% \bibliography{ZeemanPaper}% Produces the bibliography via BibTeX (use for all but very last version; in very last version: paste bbl file into this document)


\end{document}